\documentclass[conference]{IEEEtran}
\IEEEoverridecommandlockouts
\usepackage{cite}
\usepackage{amsmath,amssymb,amsfonts}
\usepackage{algpseudocode}
\usepackage{graphicx}
\usepackage[utf8]{inputenc}
\usepackage{siunitx}
\usepackage{booktabs, makecell}
\ifCLASSOPTIONcompsoc
    \usepackage[caption=false, font=normalsize, labelfont=sf, textfont=sf]{subfig}
\else
\usepackage[caption=false, font=footnotesize]{subfig}
\fi
\usepackage{nomencl}
\usepackage{multirow,tabularx}
\usepackage{etoolbox}
\usepackage{textcomp}
\usepackage{xcolor}
\usepackage{verbatim}
\usepackage{mathtools}
\usepackage{mdwtab}
\usepackage{array}

\makeatletter
\newcommand*{\rom}[1]{\expandafter\@slowromancap\romannumeral #1@}
\makeatother

\begin{document}

\title{An ICA-Based HVAC Load Disaggregation Method Using Smart Meter Data}

\author{\IEEEauthorblockN{Hyeonjin Kim, Kai Ye,\\ Han Pyo Lee, Rongxing Hu, Ning Lu}
\IEEEauthorblockA{\textit{North Carolina State University} \\
\textit{Raleigh, NC 27606, USA}\\
\{hkim66, kye3, hlee39, hru5, nlu2\}@ncsu.edu}
\and
\IEEEauthorblockN{Di Wu}
\IEEEauthorblockA{\textit{Pacific Northwest National Laboratory}\\
\textit{Richland, WA 99352, USA} \\
di.wu@pnnl.gov}
\and
\IEEEauthorblockN{PJ Rehm}
\IEEEauthorblockA{\textit{ElectriCities of North Carolina Inc.}\\
\textit{Raleigh, NC 27604, USA} \\
prehm@electricities.org}
}

\maketitle

\begin{abstract}
This paper presents an independent component analysis (ICA) based unsupervised-learning method for heat, ventilation, and air-conditioning (HVAC) load disaggregation using low-resolution (e.g., 15 minutes) smart meter data.
We first demonstrate that  electricity consumption profiles on mild-temperature days can be used to estimate the non-HVAC base load on hot days. A residual load profile can then be calculated by subtracting the mild-day load profile from the hot-day load profile. The residual load profiles are processed using ICA for HVAC load extraction. An optimization-based algorithm is proposed for post-adjustment of the ICA results, considering two bounding factors for enhancing the robustness of the ICA algorithm. First, we use the hourly HVAC energy bounds computed based on the relationship between HVAC load and temperature to remove unrealistic HVAC load spikes. Second, we exploit the dependency between the daily nocturnal and diurnal loads extracted from historical meter data to smooth the base load profile. Pecan Street data with sub-metered HVAC data were used to test and validate the proposed methods.Simulation results demonstrated that the proposed method is computationally efficient and robust across multiple customers.

\end{abstract}


\section{Introduction}
Load disaggregation is an important technique in distribution system analysis. Its results can be used in many downstream tasks, for example, customer segmentation, resource identification, and rate recommendation. When the sampling rate of available data sets is less than 1-minute, non-intrusive load monitoring (NILM) methods are often used to disaggregate the electricity consumption curves of different appliances. There are three popular data sets used for developing the NILM algorithms: UK-DALE, REDD, and REFIT, the sampling rates of which are 1, 3, and 8 seconds, respectively.  However, in practice, inputs to many load disaggregation algorithms are 15-minute smart meter data and hourly weather data. This makes those NILM methods relying on second-level meter data inapplicable.
%~\cite{liu2021samnet}~\cite{shi2019nonintrusive}.

In recent years, \textit{supervised} and \textit{unsupervised} learning methods are increasingly used for solving load disaggregation problems.
%State-of-the-art \textit{supervised learning-based} approaches are converging to a deep learning-based framework. For example, Kaselimi \emph{et al.} introduced the bidirectional long-term short memory model \cite{kaselimi2020context} and  Qian \emph{et al.} introduced the temporal convolution network (TCN) \cite{qian2021improved}. Then, transfer learning is used to transfer latent features trained by the complex appliance to simple appliances, as presented by D’Incecco \emph{et al.} in \cite{d2019transfer}.
The main disadvantage of \textit{supervised learning-based} methods is that labeled data sets are hard to obtain because sub-metered, device-level load profiles are rarely available. % to utility engineers.
Furthermore, generalization of learned knowledge from one distribution system to another is also difficult because of different weather patterns and customer use patterns. Therefore, in this paper, we choose to use the \textit{unsupervised learning-based} approach.

The majority of \textit{unsupervised learning-based} approaches are based on the temporal graphical model, especially, hidden Markov model (HMM)~\cite{bonfigli2017non,kong2016hierarchical,yan2022efhmm}. Appliances are modeled as HMM and the objective is to estimate the most likely states of appliances.
However, the main drawback of this method is that problem complexity exponentially increases with the number of appliances.
Furthermore, this method is dependent on metering data~\cite{yang2019semisupervised} with a sampling rate faster than typical smart meter data (i.e., 15-minute).

To overcome the aforementioned challenges, we propose an independent component analysis (ICA) based unsupervised load disaggregation method.
ICA is a widely used unsupervised method for blind source separation in speech recognition for speaker identification~\cite{comon1994independent}.
%For HVAC load disaggregation application,
It solely exploits information on each customer and is independent of data resolution.
In \cite{zhu2014load}, ICA was used for heat, ventilation, and air conditioning (HVAC) load disaggregation where smart meter data and wavelet analysis results are used as inputs to represent different frequency components.  However, besides HVAC loads, many other appliances also exhibit cyclic characteristics (e.g., refrigerators). This deteriorates the performance of ICA when being used in load disaggregation applications.

Our first contribution is the use of residual load profiles instead of the total load profiles as the inputs to the ICA algorithm. In~\cite{liang2019hvac}, we show that the mild-day load profile is similar to the base load profile on a hot day. Based on this finding, residual load profiles can be computed by subtracting mild-day load profiles from hot-day load profiles. Because the mild-day load profile includes weather insensitive cyclic loads, the major cyclic component left in the residual load profile is the HVAC load. This significantly improves the ICA disaggregation accuracy.

Our second contribution is the development of  an optimization-based, post-adjustment algorithm for post-processing ICA results to further improve disaggregation accuracy. The algorithm considers two bounding factors for enhancing the robustness of the ICA algorithm. First, we use the hourly HVAC energy bounds computed from the relationship between HVAC load and temperature to remove unrealistic HVAC load spikes. Second, we exploit the dependency between the daily nocturnal and diurnal loads extracted from historical meter data to smooth the base load profile.
We used the 1-minute (down-sampled to 15-minute data) Pecan Street dataset~\cite{pecan}, where sub-metered HVAC data are available for evaluating the performance of the algorithm.

\section{Methodology} \label{Methodology}

\begin{figure}[t!]
	\centering
	\vspace{-.3cm}
	\includegraphics[width=2.3in]{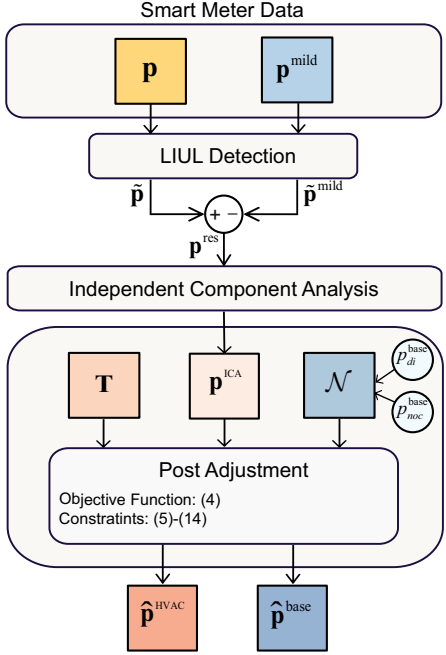}
	\caption{Workflow of the HVAC load disaggregation process.}
\label{fig:workflow} % Give a unique label
\vspace{-.1cm}
\end{figure}

This section presents the overall modeling framework, the computing of the residual load profile, and the ICA and the post-processing algorithms.
%, a few key findings from the smart meter data of individual customers are represented. In addition, the proposed HVAC disaggregation model is introduced.

\subsection{Load Disaggregation Work Flow} \label{workflow}
Let $\mathbf{P}$ be the daily load profile matrix and $p_{i,j}$ represent the $i^{\mathrm{th}}$ ($i \in N$) data sample of the $j^{\mathrm{th}}$ ($j \in M$) day, where $N$ and $M$ is the number of data samples in a day and the number of hot days, respectively.
Let $\mathbf{P}^{\mathrm{mild}}$ be the mild-day load profile matrix and $p^{\mathrm{mild}} _{i,k}$ represent the $i^{\mathrm{th}}$ data sample of the $k^{\mathrm{th}}$ ($k \in K$) day, where $K$ is the number of mild days. For 15-minute smart meters data set, $N=96$. Thus, $p_{1:N,j}$ is the daily load profile for the $j^{\mathrm{th}}$ hot day and $p^{\mathrm{mild}} _{1:N,k}$ is the $k^{\mathrm{th}}$ mild day.

As shown in Fig.~\ref{fig:workflow}, we first remove infrequently used loads (LIUL) (e.g., dryer and electric hot water heater) from each daily profile in $\mathbf{P}$ and  $\mathbf{P}^{\mathrm{mild}}$ using the LIUL filtering algorithm introduced in~\cite{liang2019hvac}.
The load profile matrices with LIUL removed are represented by  $\tilde{{\mathbf{P}}}$ and $\tilde{{\mathbf{P}}}^{\mathrm{mild}}$, respectively.
Next, we calculate the residual load profile matrix, ${\mathbf{P}}^{{\text{res}}}$, by subtracting $\tilde{{\mathbf{P}}}^{\mathrm{mild}}$ from $\tilde{{\mathbf{P}}}$ (see Section~\ref{mildday}). Then,  ${\mathbf{P}}^{{\text{res}}}$ is fed to the ICA algorithm (see Section~\ref{ICA}) for extracting the HVAC load profile matrix, ${\mathbf{P}}^{{\text{HVAC}}}$.
Finally, a residual minimization fine-tuning algorithm (see Section \ref{postprocessing}) is used to adjust $\mathbf{P}^{\text{HVAC}}$ and $\mathbf{P}^{\text{base}}$ to further improve the disaggregation accuracy.

\subsection{Classification of Hot and Mild Days} \label{mildday}

We define a mild day as a day with no heating or cooling loads. In addition, we define a hot day as a day with HVAC loads. The method for hot and mild day classification introduced in~\cite{liang2019hvac} is based on temperature thresholds. However, some customers will turn on their HVAC even in a mild day, causing some resultant mild days to contain a small amount of HVAC loads. This can significantly reduce the accuracy of HVAC load disaggregation in later steps.  Therefore, we improve the method by adding a verification step: power distribution comparison.

\begin{figure}[th]
\centering
\vspace{-.3cm}
\subfloat[]{\includegraphics[width=1.5in]{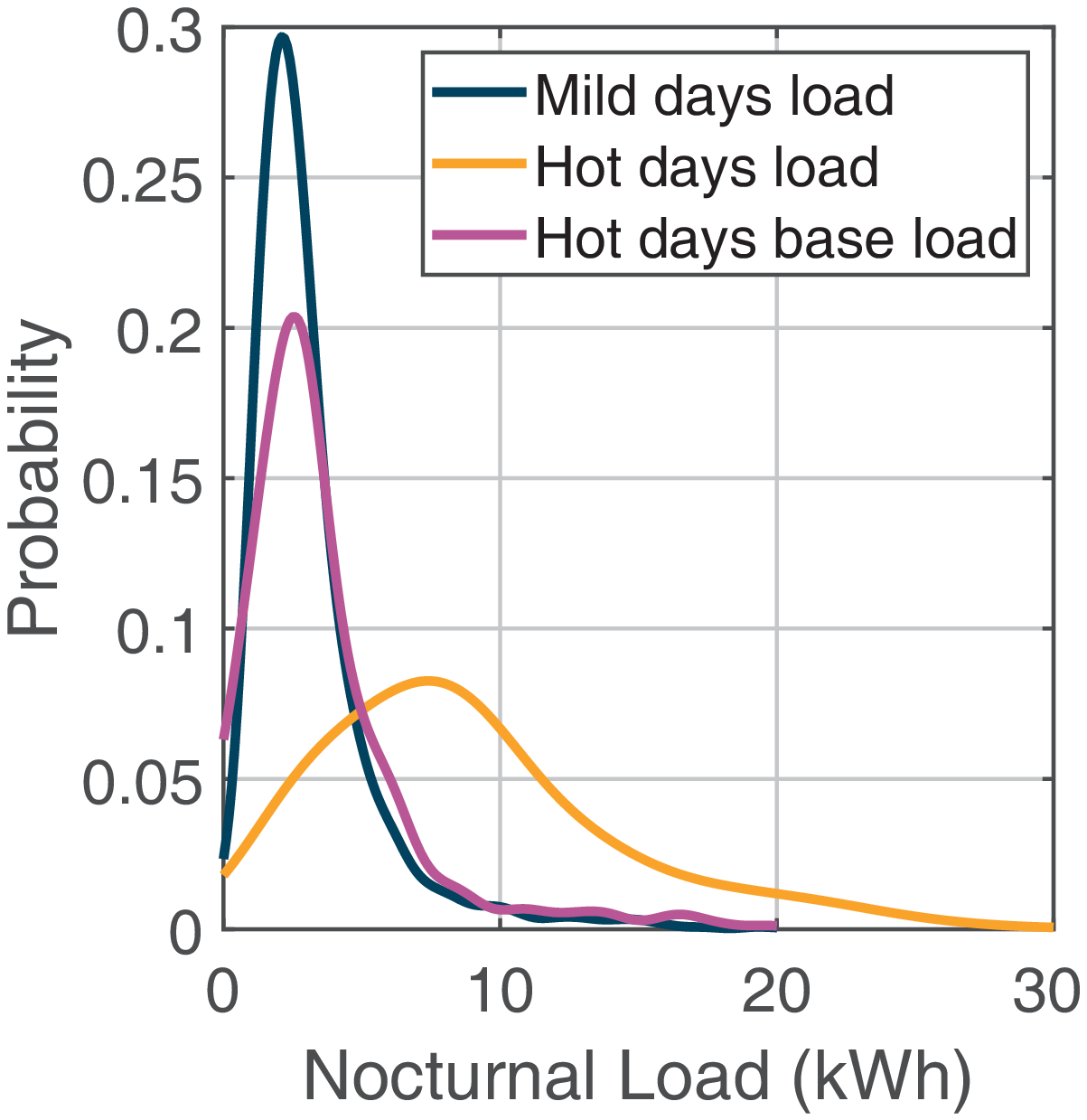}\label{noc}}
\subfloat[]{\includegraphics[width=1.5in]{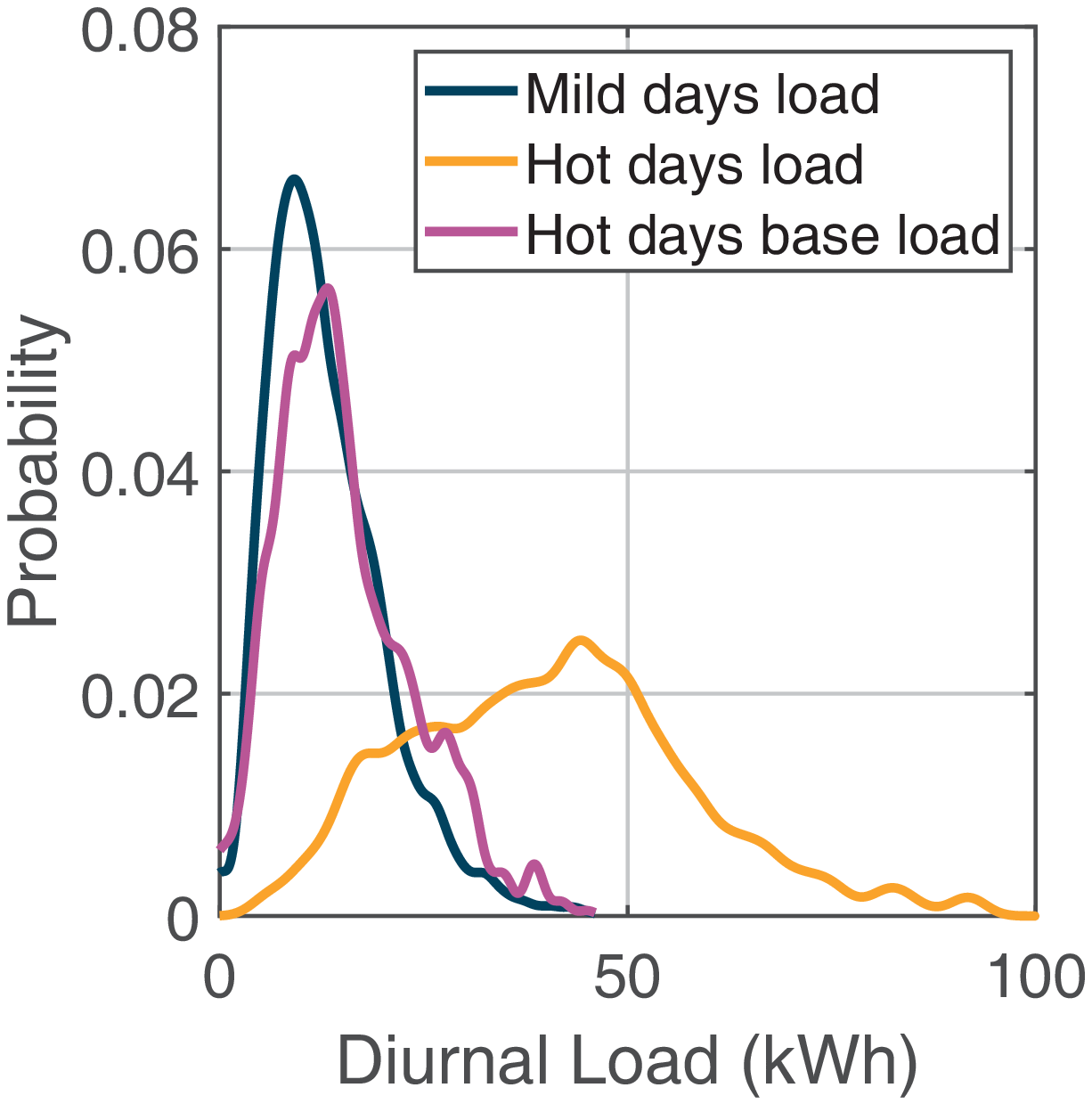}\label{dir}}\\
\caption{Comparison of power consumption distributions: base loads in hot days, mild-day loads, and hot-day loads, respectively.}
\label{ComparisonPDF}
\vspace{-.2cm}
\end{figure}

As shown in Fig.~\ref{ComparisonPDF}, the distribution of the nocturnal and diurnal base loads in hot days overlap well with the distribution of the loads for mild days. Therefore, by comparing the power distribution of load profiles, we can separate the mild days from the hot days by comparing its power distribution.

\begin{figure}[th]
	\centering
	\vspace{-.3cm}
	\includegraphics[width=3.0in]{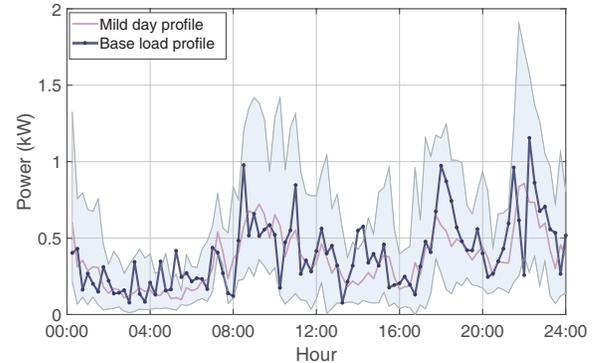}
	\caption{Average load profiles for the actually measured base load (the solid blue line) and the calculated base load using  mild-day load profiles (the solid red line). The shaded area shows the $\pm$ one standard deviation against the calculated base load load profile.}
\label{LoadProfile} % Give a unique label
\vspace{-.1cm}
\end{figure}

As shown in Fig.~\ref{LoadProfile}, the base load profiles, $\mathbf{P}_j^{\rm{base}}$, in hot days are bounded by the mild day profiles, $\mathbf{P}^{\rm{mild}}$, so we have
\begin{equation}
    \mathbf{P}_j^{\rm{base}} = {\tilde{\mathbf{P}}}^{\rm{mild}} \cdot {\boldsymbol{\beta }}+ {\boldsymbol{\theta}}^b\\
\end{equation}
where ${\boldsymbol{\theta }}^b$ is an adjustment vector with $N \times 1$ dimension and $\boldsymbol{\beta }$ is a multiplier.

By subtracting $K$ mild day load profiles from the $j^{\mathrm{th}}$ day load profile one by one, we can obtain an ensemble of residual load profiles, represented by ${\mathbf{P}}^{{\text{res}}}_j$, which is an $N \times K$ matrix. Each column of ${\mathbf{P}}^{{\text{res}}}_j$ contains the actual HVAC load profile, $\mathbf{p}^{\text{HVAC}}_{j}$, plus a random power variation, $\epsilon_j$, both of which are $N \times 1$ vector, so we have
\begin{equation} \label{eqn:ICA}
{\mathbf{P}}_j^{{\text{res}}} = [\begin{array}{*{20}{c}}
  {{\mathbf{p}}_j^{{\text{HVAC}}}}&{\epsilon}_j
\end{array}] \cdot {\mathbf{A}}
\end{equation}
where $\textbf{A}$ is an $2 \times K$ mixing matrix.

%Load profiles of each customer on mild temperature days can approximate base load profiles, which are without HVAC load consumption. This is based on the assumption that the consumption pattern of each customer generally has its own tendencies. In Fig.~\ref{Scatter1}, three different distributions of load are represented. As described in former studies, analyzing the diurnal and nocturnal load of each customer can help derive characteristics of load profiles~\cite{bu2021disaggregating}. It can be found that hot days loads show different consumption patterns due to HVAC load usage: bigger consumption, larger variability. However, the distribution of base load on hot days and mild days load overlap in most of the area. This implies that exploiting the known distribution of mild days load of each customer can help illustrate the unknown distribution of base load in target days.

\subsection{Independent Component Analysis} \label{ICA}

ICA is a widely used unsupervised method for blind source separation in signal processing ~\cite{hyvarinen2000independent}. ICA has also been used in NILM for retrieving unlabeled sources.
%For example, in ~\cite{rahimpour2017non}, Rahimpour \emph{et al.} use ICA to decompose the total feeder level load into commercial and industrial loads \textcolor{red}{using wavelet analysis}.
To extract ${\bf{P}}_j^{\rm{ICA}}$ from ${\bf{P}}_j^{\rm{res}}$, we formulated the ICA problem as
%is used for the cases where multiple unobservable blind sources $\textbf{s}$ are mixed in the observable variable $\textbf{x}$ represented as:
\begin{equation} \label{ICAInverse}
   {\bf{P}}_j^{\rm{ICA}} =  {\bf{P}}_j^{\rm{res}}\textbf{W}
\end{equation}
where $\textbf{W}$ is an $K \times 2$ unmixing matrix (the pseudo inverse of $\textbf{A}$) and ${\bf{P}}_j^{\rm{ICA}}$ is the calculated HVAC load profile for the  $j^{\mathrm{th}}$ day. Note that the residual load profiles are obtained by subtracting the total load profile of the $j^{\mathrm{th}}$ day with $K$ different mild day load profiles, which we proved to bear the similar load shapes as the base load profile of the $j^{\mathrm{th}}$ day. Thus,
the ICA algorithm only needs to identify a high percentage, cyclic HVAC load profile and a low percentage, highly random end use load consumption from the residual load profiles. This significantly improves the identification accuracy.

%Assuming that each source in $\textbf{s}$ are independent, joint probablitiy distribution of $\textbf{s}$ can be formulated as a product of marginals. Then, by applying the relationship in (1) joint probability of $\textbf{s}$ can be converted as the function of $\textbf{x}$. Using maximum likelihood estimation, optimal value of $\textbf{W}$ can be derived. Detailed algorithmic explanation can be found in.

% However, designing matrix $\textbf{x}$ and $\textbf{s}$ is different for each study. For example, in~\cite{zhu2014load}, total feeder level load and decomposed signals of total load by wavelet analysis are used to decompose total load into commercial and industrial loads. For our cases we use residual profile as a $\textbf{x}$ which is represented as below:
\begin{figure}[th]
\centering
\vspace{-.3cm}
\includegraphics[width=3 in]{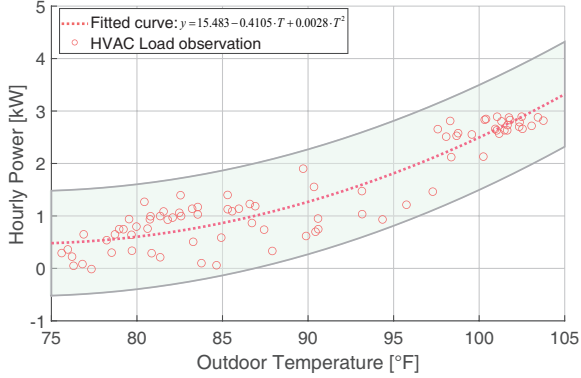}
\caption{HVAC hourly power consumption with respect to different outdoor temperatures.}
\label{Temp}
\vspace{-.1cm}
\end{figure}

\subsection{Fine-tuning of the ICA Results} \label{postprocessing}
The estimated HVAC and base load profiles obtained by ICA can be fine-tuned to further improve the load disaggregation accuracy. Note that because we will estimate the HVAC load and the base load for the $j^{\mathrm{th}}$ day, the index $j$ is omitted for all variables.

Let $\hat{{\bf{p}}}^{{\rm{HVAC}}}$ and $\hat{\bf{p}}^\text{base}$ be the final estimation of the HVAC load and the base load of hot days, respectively, the objective function of the fine-tuning process can be formulated as

\begin{multline}
\label{QGB}
\min_{\bf{p},\boldsymbol{\alpha},\boldsymbol{\beta},\boldsymbol{\gamma},\boldsymbol{\theta}}
L_{\mathrm{shape}} +
\lambda_1\left\|\boldsymbol{\theta}^h \right\|_2^2
+\lambda_2\left\|\boldsymbol{\theta}^b\right\|_2^2
-\lambda_3L_{\mathrm{PDF}}
\end{multline}
subject to:
\begin{align}
& L_{\mathrm{shape}}  = \left\| \bf{p} - \hat{\bf{p}}^{\rm{HVAC}} - \hat{\bf{p}}^\text{base} \right\|_2^2 \\
& L_{\mathrm{PDF}}= D_{\text{KL}}({\mathcal N}(\hat{p}_{di}^\text{base},\hat{p}_{noc}^\text{base})||{\mathcal N}( \tilde{p}_{\mathrm{di}}^\text{mild},\tilde{p}_{\mathrm{noc}}^\text{mild})) \\
&\hat{{\bf{p}}}^{{\rm{HVAC}}}= \alpha  \cdot {{\bf{p}}}^{{\rm{ICA}}} + {\boldsymbol{\theta}}^h\\
&\hat{{\bf{p}}}^{{\rm{base}}} = {\tilde{\bf{P}}}^{{\rm{mild}}} \cdot {\boldsymbol{\beta }}+ {\boldsymbol{\theta }}^b\\
&\hat{{\bf{p}}}_{{\rm{hour}}}^{{\rm{HVAC}}} = {{\bf{\gamma }}_1} \cdot {\bf{T}} + {{\bf{\gamma }}_2} \cdot {\bf{T}} \odot {\bf{T}}\\
&\left| \frac{1}{4} \cdot {\sum\limits_{i = 4k-3}^{4k} {{\hat{p}_{i}^{\text{HVAC}}}}  - \hat{{p}}_{{\rm{hour}}, k}^{{\rm{HVAC}}}} \right| \le \varepsilon\,\,\,\,\,\forall k \in \{1, 2, ...,24\}\\
&\hat{p}_{di}^{\rm{base}} = \frac{1}{4} \cdot\sum\limits_{i \in N_{di}}^{} \hat{{{p}}}^{{\rm{base}}}_{i}\\
&\hat{p}_{noc}^{\rm{base}}  = \frac{1}{4} \cdot\sum\limits_{i \in N_{noc}}^{} \hat{{{p}}}^{{\rm{base}}}_{i}\\
&0 \le \hat{{p}}_{i}^{{\rm{HVAC}}} \le {p_{i}}\,\,\,\,\,\forall i \in N\\
&0 \le \hat{{p}}_{i}^{{\rm{base}}} \le {p_{i}}\,\,\,\,\,\forall i \in N
\end{align}
where $p_{di}^{\rm{base}}$ and $p_{noc}^{\rm{base}}$ represent the nocturnal and diurnal base loads, respectively, $\hat{{\bf{p}}}_{\rm{hour}}^{\rm{HVAC}}$ is the hourly HVAC load, $\bf{T}$ is a vector of outdoor temperature, $\gamma_1$ and $\gamma_2$ are coefficients, $\theta^h$ and $\theta_b$ are adjustment vectors, and $\varepsilon$ is an error margin.

The first loss term calculated by (5) calculates the point-to-point load disaggregation error. The second loss term calculated by (6) represents the similarity between diurnal and nocturnal power consumption distributions of the final base load and the mild-day-computed base load.  To compensate non-linearity,  the final estimation of the HVAC and the base load, $\hat{{\bf{p}}}^{{\rm{HVAC}}}$ and $\hat{\bf{p}}^\text{base}$, are scaled by $\alpha$ and $\beta$ and adjusted by $\theta_h$ and  $\theta_b$ by (7) and (8), respectively. 
%We use (13)-(14) to ensure that the estimated HVAC and base-load hourly energy consumption are lower than or equal to the actual load profile.  
(13) and (14) ensure that, at each timestamp, $\hat{{\bf{p}}}^{{\rm{HVAC}}}$ and $\hat{\bf{p}}^\text{base}$ are both lower than the total load.

As shown in Fig.~\ref{Temp}, HVAC load can be assumed to have quadratic relationship with outdoor temperature. To incorporate hourly temperature into the estimation, we first calculate the hourly HVAC load, $\hat{{\bf{p}}}_{\rm{hour}}^{\rm{HVAC}}$, by (9). 
Then, the sum of the estimated HVAC load in each hour is bounded by $\hat{{\bf{p}}}_{{\rm{hour}}}^{{\rm{HVAC}}}$ with a margin of $\varepsilon$ in (10).

\begin{figure}[t!]
\centering
\includegraphics[width=3.4in]{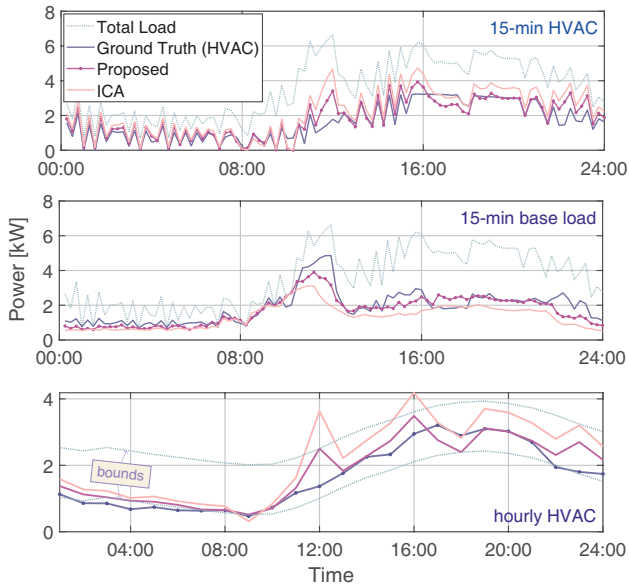}
\caption{Load disaggregation results for one house in a summer day}
\label{Result1}

\end{figure}

\section{Simulation Results}
The simulation results and performance evaluation are presented in this section.
\subsection{Simulation Setup and Performance Evaluation Metrics}
Pecan Street data sets (including 150 total household loads and the corresponding sub-metered HVAC loads) are used to develop and validate the proposed algorithm. The original data length is 1-year with 1-minute data resolution. We down-sample the 1-minute data to 15-minute. The testing data sets are constructed by randomly selected 30 days in summer from 70 customers in Austin, Texas.

To evaluate the fine-tuning algorithm performance, we set up three cases. The first case is without a log-likelihood term, i.e., (6). The second considers the log-likelihood with the distribution obtained from a single user. The third case considers the log-likelihood obtained from multiple users.

We compare our method with two benchmark methods. The first method (introduced in [8]) is to compute the average of all mild-day load profiles and use it as the base load profile, which can then be subtracted from the total load profile to get the HVAC load profile.  The second method is the ICA method without fine-tuning.

%-based model outputs base load estimation as the average of all mild-days load profiles.

The disaggregation accuracy is evaluated using the normalized mean absolute error ($n$MAE) and the normalized energy error ($n$EE), computed by

\begin{align}
&n\text{MAE}=\frac{1}{M} \cdot \sum\limits_{j = 1}^M \sum\limits_{i = 1}^N {\frac{{\left| {\bf{P}^{\mathrm{*}}_{i,j} - {\bf{P}_{i,j}}} \right|}}{\text{Rating}}}\\
&n\text{EE} = \frac{{\left| {\sum\limits_{j = 1}^M \sum\limits_{i = 1}^N{p_{i,j}^*}  - \sum\limits_{j = 1}^M \sum\limits_{i = 1}^N{p_{i,j}}}  \right|}}{{\sum\limits_{j = 1}^M \sum\limits_{i = 1}^N{p_{i,j}}}}
\end{align}
where the variable with `$*$' represents the estimated value.

\subsection{Efficacy of Fine-tuning}
\begin{figure}[t!]
	\centering
	\includegraphics[width=3.2in]{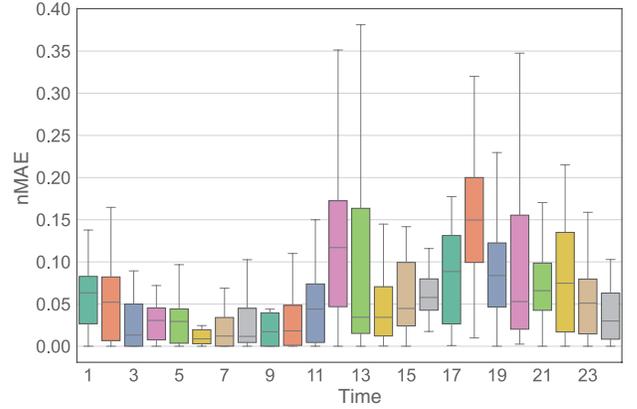}
	\caption{Hourly box plots of load disaggregation errors for summer load profiles (with fine-tuning). The total data length is 30 days.}
\label{fig6} % Give a unique label

\end{figure}

In this subsection, we compare two cases: with and without fine-tuning. The results are shown in Figs.~\ref{Result1} and \ref{fig6}, from which, the following observations can be made:

\begin{itemize}
    \item Without fine-tuning, the HVAC (subplot 1) and base load (subplot 2) profiles obtained by the ICA method has large overshoots or undershoots compared with the with fine-tuning case. This is because the ICA results are not bounded.
    \item As shown in subplot 3, the fine-tuning algorithm uses the temperature-power relationship to derive upper and lower hourly HVAC consumption bounds. This effectively reduces the occurrence of over- and under-shoots, which, consequently, drastically reduces the point-to-point mismatch.
    \item In Fig.~\ref{fig6}, we show the disaggregation error distribution for each hour of the day of a randomly selected load profile, the accuracy of which lies in the middle of all 70 load profiles. The results show that the performance is the best in the early morning and the worst during the lunch and dinner times. This is because there are multiple appliances in use and the cycling of the HVAC unit may be affected by internal heating sources  (e.g., heat from cooking) besides the outdoor temperature. The highest error happens at 18:00 with a median $n$MAE of 0.15.  However, most of the demand response programs target at the time period from 15:00 to 17:00, during which, the overall performance of the algorithm is satisfactory. 

\end{itemize}

%\subsection{Hourly disaggregation performance}

%, disaggregation results of a single customer on a single day in summer are represented. In Fig.~\ref{R1} and Fig.~\ref{R2}, disaggregation results on HVAC load and base load are represented by 15-minute granularity. It is shown that both the proposed and ICA method can well detect ramping events induced by HVAC load consumption in most of the period. However, in some time stamps, the HVAC load estimation of the ICA method can overshoot as it is not bounded due to the absence of information on the HVAC rating of each customer. Leveraging hourly temperature can help this issue by letting them serve as upper and lower bounds in optimization formulation. Therefore, in Fig.~\ref{R3}, hourly HVAC load baseline is represented as green lines, which are derived by daily the temperature profile. As a result bounding the decision variable $\bf{p}_h$ by using temperature information helps to remove overshooting issues in preliminary estimation.

\subsection{Efficacy of Matching Base Load Distribution}
\begin{figure}[h]
	\centering
	\includegraphics[width=3.6in]{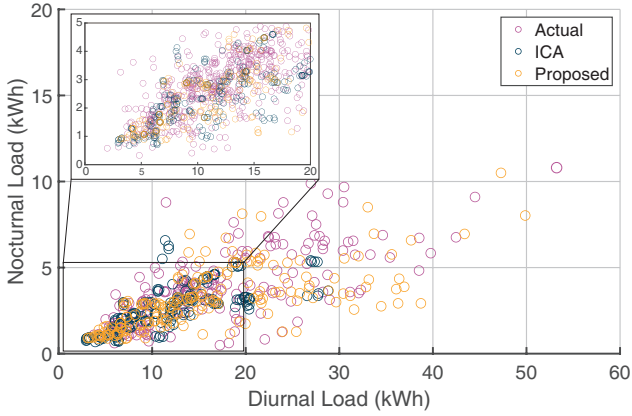}
	\caption{Bivariate distributions of the diurnal and nocturnal base loads in the target days}
\label{fig7} % Give a unique label

\end{figure}

In this subsection, the effect of including constraint (6) in the fine-tuning process (4) is evaluated. As shown in Fig.~\ref{fig7},  
%shows three different bivariate distributions of the diurnal and nocturnal base loads in the target days. The results show that the distribution of 
the ICA identified base load has a larger mismatch than the mild day computed base load (marked as the "proposed" in Fig. 7) when comparing with the actual distribution. This is further proved by the results shown in Table~\ref{t2}.
%, where the distance between parameters of the proposed and the actual base loads are indeed closer than that between the ICA calculated base load and the actual base load. 
Thus, in the fine-tuning, we can use (6) to correct the ICA calculated base load profile to achieve a more accurate point-to-point match in base load disaggregation.

\subsection{Comparing with the Benchmark Methods}
In Table~\ref{t1}, the numerical performance of proposed models and benchmark models are compared. It is shown that by adding the fine-tuning as the post-adjustment process, the point-to-point estimation error can be reduced for all three cases compared to benchmark models. However, the accumulated error of the proposed model represented by $n$EE are about the same as the ICA model without fine-tuning. This is because the post-adjustment process adjusted the ICA result up and down to smooth the curve, making the accumulated energy consumption of the HVAC load remain approximately the same. Finally, the standard deviation of $n$MAE for the proposed model is smaller than that of the benchmark model, showing a more robust and consistent estimation performance. 

\begin{figure}[t!]
\centering
\subfloat[Proposed]{\includegraphics[width=3in]{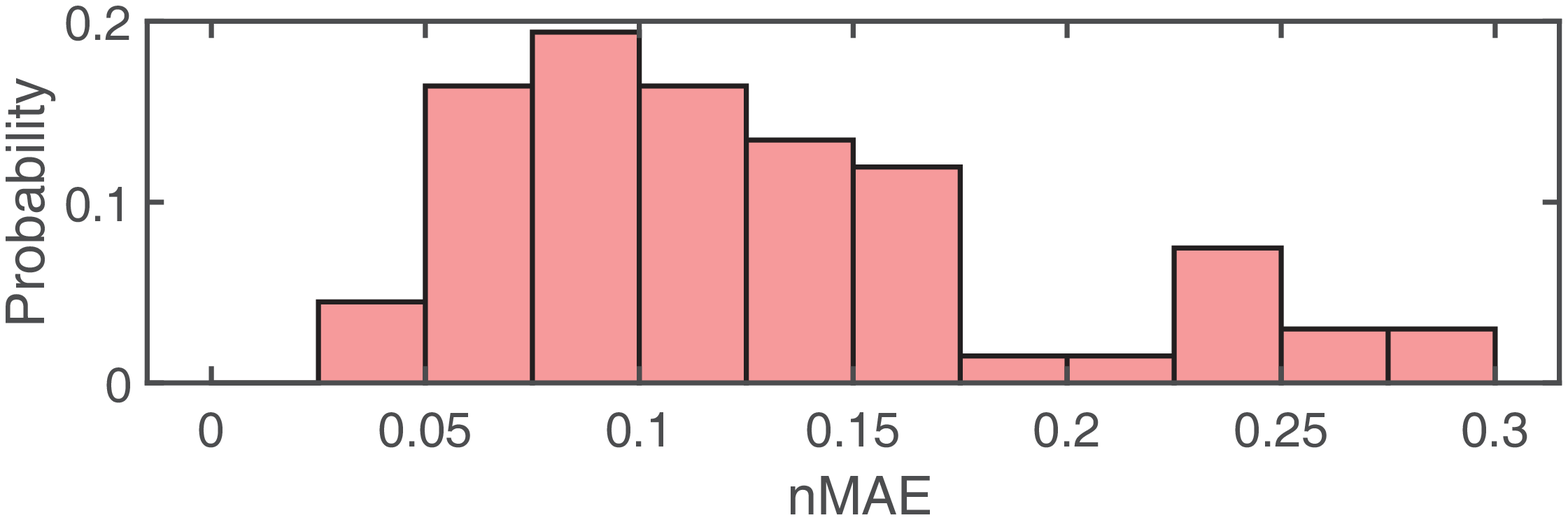}\label{sk1}}\\
\subfloat[ICA]{\includegraphics[width=3in]{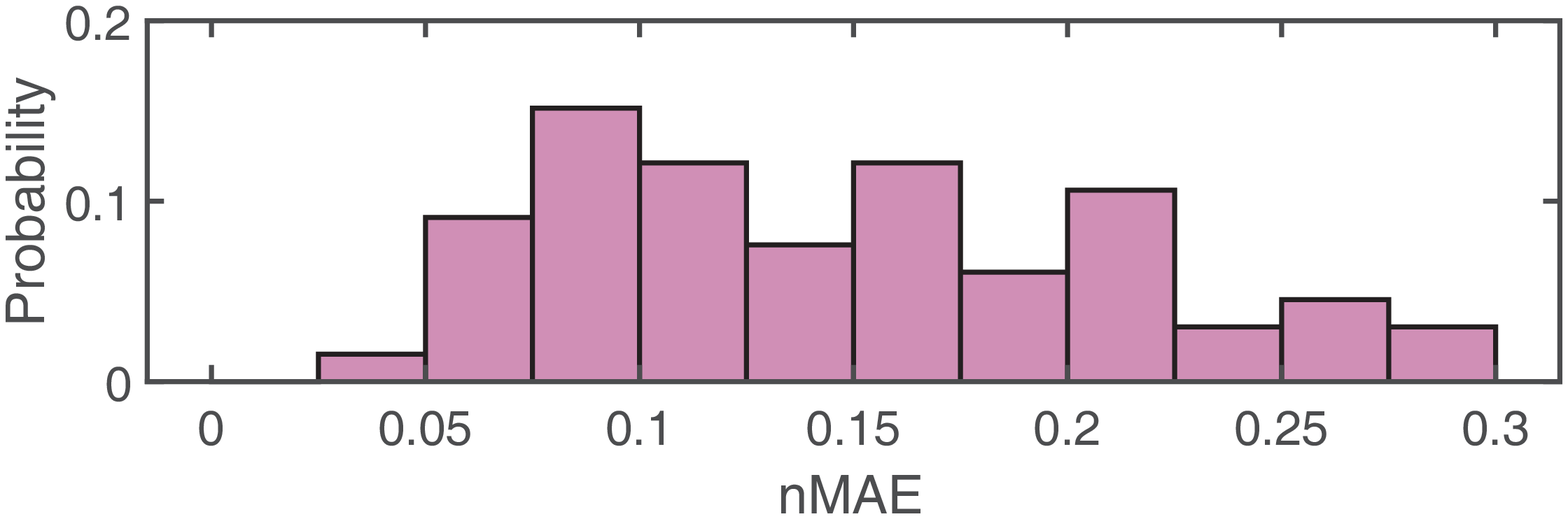}\label{sk1}}\\

\caption{Empirical distributions of nMAE of 70 customers }
\label{Emp}
\end{figure}

This trend can also be found in Fig.~\ref{Emp}, where the empirical distributions of $n$MAE are shown. Note that each error observation is collected by averaging the resultant errors of each customer on multiple days to show the robustness of model across multiple customers. While $n$MAE of the ICA method without fine-tuning are distributed almost uniformly from 0.05 to 0.3, the error distribution of Case III of the proposed model is centered around 0.1.

\begin{table}[t!]
	\begin{center}
		\caption{Mean Values and Standard Deviations of the Distribution of the Base Load 15-minute Power Consumption}
		\label{t2}
		\begin{tabular}{cccc}
		\toprule
			\toprule % <-- Toprule here
\scalebox{0.88} {Parameters} & \scalebox{0.88} {\textbf{Actual}} & \scalebox{0.88}{\textbf{ICA}}  &\scalebox{0.88} { \textbf{Proposed}}\\
\midrule

 $\boldsymbol{\mu}$ & $\left[\begin{array}{*{20}{c}}
 \scalebox{0.88} {15.40} \\
  \scalebox{0.88} {3.16}
\end{array}\right]$ & $\left[\begin{array}{*{20}{c}}
  \scalebox{0.88} {12.12} \\
  \scalebox{0.88} {2.69}
\end{array}\right]$ & $\left[\begin{array}{*{20}{c}}
   \scalebox{0.88} {16.45} \\
  \scalebox{0.88} {3.22}
\end{array}\right]$\\
\midrule
 $\boldsymbol{\Sigma}$ & $\left[\begin{array}{*{20}{c}}
  \scalebox{0.88} {79.99}&\scalebox{0.88}{13.12} \\
  \scalebox{0.88}{13.12}&\scalebox{0.88}{8.02}
\end{array} \right]$ & $\left[\begin{array}{*{20}{c}}
  \scalebox{0.88}{32.13}&\scalebox{0.88}{5.08} \\
  \scalebox{0.88}{5.08}&\scalebox{0.88}{1.76}
\end{array} \right]$ & $\left[\begin{array}{*{20}{c}}
  \scalebox{0.88}{91.30}&\scalebox{0.88}{10.56} \\
  \scalebox{0.88}{10.56}&\scalebox{0.88}{3.75}
\end{array}\right]$ \\

		    \bottomrule
			\bottomrule % <-- Bottomrule here
		\end{tabular}
	\end{center}

\end{table}

\begin{table}[t!]
	\begin{center}
		\caption{Performance Comparison for Different Load Disaggregation Methods}
		\label{t1}
		\begin{tabular}{cccccccc}
		\toprule
			\toprule % <-- Toprule here
&\multicolumn{2}{c}{\textbf{Benchmark Models}} & & \multicolumn{3}{c}{\textbf{Proposed Model}}\\
 \cmidrule{2-3}  \cmidrule{5-7}
&\textbf{Average} & \textbf{ICA} & & \textbf{Case 1} & \textbf{Case 2} & \textbf{Case 3}\\
\midrule
$n$MAE (\%) & 15.48 & 14.80 & & 13.75 & 13.87 & 12.92\\
$n$EE (\%) & 16.35 & 15.82 & &14.07 & 15.73& 14.21\\
std($n$MAE) & 7.52 & 9.32 & & 6.09 & 7.42 & 5.87\\

		    \bottomrule
			\bottomrule % <-- Bottomrule here
		\end{tabular}
	\end{center}

\end{table}

\section{Conclusion}
In this paper, we present an ICA-based HVAC load disaggregation model enhanced by a fine-tuning process.  This approach enables the HVAC load disaggregation using low-resolution smart meter data as inputs.  We first proved that by using the mild day load profiles, one can successfully extract the base load profile and approximate the distribution of the base load electricity consumption. Then, we show that the ICA method can be used more effectively on the residual load profiles (computed by subtracting the base load from the hot day load profiles) instead of on the hot day load profiles directly. Next, through fine-tuning, the preliminary results obtained by the ICA method can be further improved. The proposed method is compared with benchmark models and shows consistent performance across multiple customers.  Our future work is to apply the method to disaggregate load profiles with behind-the-meter solar photovoltaic systems and electric vehicle charging loads.
\bibliographystyle{IEEEtran}
\bibliography{conf}

\end{document}